\documentclass[conference,letterpaper]{IEEEtran}

\usepackage[utf8]{inputenc} 
\usepackage[T1]{fontenc}
\usepackage{url}
\usepackage{amssymb}
\usepackage{ifthen}
\usepackage{cite}
\usepackage[cmex10]{amsmath} 
\newcommand\numeq[2]%
  {\stackrel{#1}{#2}}
  
\usepackage{epsfig,rotating,setspace,latexsym,amsmath,epsf,amssymb,amsfonts,bm,theorem,cite,algorithm,graphicx,epsf,authblk,epstopdf,color,algpseudocode,bbm,subcaption}

\newtheorem{theorem}{Theorem}
\newtheorem{lemma}{Lemma}
\newenvironment{Proof}[1]{\medskip\par\noindent{\bf Proof:\,}\,#1}{{\mbox{\,$\blacksquare$}\par}}

\begin{document}

\title{Timely Updating with Intermittent Energy and Data for Multiple Sources over Erasure Channels} 


\author{Christopher Daniel Jr.$\qquad$Ahmed Arafa\\Department of Electrical and Computer Engineering\\ University of North Carolina at Charlotte, NC 28223\\
$\quad$\emph{cdanie36@uncc.edu}$\qquad$\emph{aarafa@uncc.edu}}

\maketitle

\begin{abstract}

A status updating system is considered in which multiple data sources generate packets to be delivered to a destination through a {\it shared} energy harvesting sensor. Only one source's data, when available, can be transmitted by the sensor at a time, subject to energy availability. Transmissions are prune to erasures, and each successful transmission constitutes a status update for its corresponding source at the destination. The goal is to schedule source transmissions such that the {\it collective} long-term average {\it age-of-information} (AoI) is minimized. AoI is defined as the time elapsed since the latest successfully-received data has been generated at its source. To solve this problem, the case with a single source is first considered, with a focus on {\it threshold} waiting policies, in which the sensor attempts transmission only if the time until {\it both} energy and data are available grows above a certain threshold. The {\it distribution} of the AoI is fully characterized under such a policy. This is then used to analyze the performance of the multiple sources case under {\it maximum-age-first} scheduling, in which the sensor's resources are dedicated to the source with the maximum AoI at any given time. The achievable collective long-term average AoI is derived in closed-form. Multiple numerical evaluations are demonstrated to show how the optimal threshold value behaves as a function of the system parameters, and showcase the benefits of a threshold-based waiting policy with intermittent energy and data arrivals.


\end{abstract}

\section{Introduction}

In remote sensing applications, maintaining {\it timely} delivery of status updates at the destinations regarding the sensed environments is necessary to take informative decisions. This becomes more challenging in situations where sensors have stringent energy budgets and data storage limits, and when the sensed data is transmitted over noisy channels. Inspired by the {\it age-of-information} (AoI) metric originally introduced in~\cite{kaul_aoi} to assess the freshness of data, this paper provides solutions to these challenges for sensing settings with multiple sources.



In this work, status updating for multiple data sources using a shared energy harvesting sensor over an erasure channel is analyzed. The sources' data and the sensor's energy arrive according to Poisson processes of different rates. The sensor can only serve one data source at a time, with scheduling and transmission policies needed to be designed to optimally manage the arriving energy to transmit the arriving data. The goal is to minimize the {\it collective} long-term average AoI of all sources at the destination. AoI is defined as the time elapsed since the latest successfully-received data has been generated at its source. We analyze the benefits of idle waiting {\it after} both energy and data are available for a given source, with a focus on {\it threshold} policies, in which a new transmission occurs only if the time until energy and data arrives surpasses a certain threshold. Idle waiting before updating has been analyzed previously in \cite{zou_waiting_aoi} for a single source, yet in a non-energy-harvesting setting with fixed waiting times, and in the single-source energy harvesting work in \cite{yates_age_eh}, yet with a first-come first-serve discipline with infinite battery and data storage. Our work focuses on last-come first-serve discipline with preemption, which is better for AoI minimization.

For our setting, we provide {\it closed-form} expressions for: (1) the AoI {\it distribution}; (2) the long-term average AoI for a single source; and (3) the collective long-term average AoI for multiple sources under {\it maximum-age-first} scheduling, in which the source with the maximum AoI is given priority over others. We then show how the optimal threshold value behaves as a function of the system parameters, such as data and energy arrival rates, erasure probability, and the number of sources. 

\begin{figure}[t]
\centering
\includegraphics[scale=.525]{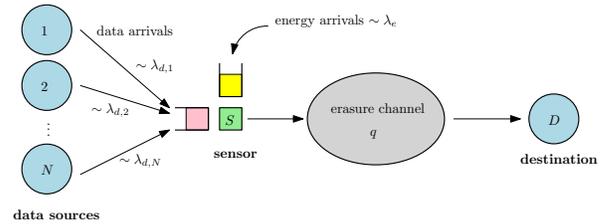}
\caption{System model overview: status updating with multiple sources using a shared energy harvesting sensor.}
\label{fig:sys-mod}
\vspace{-.25in}
\end{figure}



\noindent{\bf Related Works.} Status updating with energy harvesting sensors has been studied in, e.g., \cite{yates_age_eh, jing-age-online, jing-age-erasures-infinite-jour, shahab-age-online-rndm, farazi-aoi-eh-preempt, zheng-aoi-eh-queue, baknina-updt-info, arafa-age-online-finite, bacinoglu-aoi-eh-finite-gnrl-pnlty, arafa-age-erasure-no-fb, arafa-age-erasure-fb, krikidis-aoi-wpt, stamatakis-aoi-eh-alarm, pappas-AoI-cache, ozel-aoi-eh-sensing, chen2021aoiMAC, hirosawa2020aoiEH, jaiswal2020aoiEH, ko2020aoiEH, abdelmagid2020aoiWPT, sleem2020aoiWPT, gindullina2020aoiEH}, and can be categorized according to whether the energy harvested is known a priori (offline) or causally (online), or whether data can be generated at will or is exogenous. Our work in this paper is {\it online with exogenous data arrivals.} The works in \cite{shahab-age-online-rndm, farazi-aoi-eh-preempt} analyze AoI for a single source using tools from stochastic hybrid systems; differently, we introduce the notion of threshold waiting before updating and show that it can enhance the achievable AoI. This paper also extends the work in \cite{arafa-age-erasure-fb} to multiple data sources yet with exogenous arrivals. References \cite{hirosawa2020aoiEH, jaiswal2020aoiEH, gindullina2020aoiEH} study multiple data sources. The work in \cite{hirosawa2020aoiEH} focuses on analyzing the performances of TDMA/FDMA scheduling. Reference \cite{jaiswal2020aoiEH} follows an MDP framework in a discrete-time setting with a finite time horizon; the optimal policy is such that the sensor first probes the channel if the maximum AoI grows above a certain threshold, and then decides on sampling the source with the maximum AoI if the probed channel conditions are better than a certain threshold as well. Different from \cite{jaiswal2020aoiEH}, we consider an infinite time horizon setting, with exogenous data arrivals, and provide analytical expressions for AoI under Poisson energy arrivals. The work in \cite{gindullina2020aoiEH} considers the notion of source diversity when multiple sources monitor the same physical phenomenon with different costs.



\section{System Model and Problem Formulation}

We consider a system composed of $N$ sources of time-varying data that are to be monitored at a remote destination through the help of a {\it shared} energy harvesting sensor (transmitter). Source $j$'s data is generated in packets according to a Poisson process of rate $\lambda_{d,j}$, with each packet containing a time-stamp of its generation time. Each generated data packet is fed into the sensor's data buffer. However, the sensor is capable of only holding {\it one} data packet at a time, and it needs to decide on whether to discard newly-arriving data packets or preempt the currently-held ones, if any.

Further, the sensor relies on energy harvested from nature to communicate. Energy arrives in units according to a Poisson process of rate $\lambda_e$, with each unit capable of transmitting one data packet. The sensor is equipped with a battery of unit size to save the incoming energy. All processes (sources' data and sensor's energy) are independent. Only when both energy and data are available the sensor may transmit. Transmissions are instantaneous,\footnote{This is a reasonable approximation for transmission rates that are much larger than the data and energy arrival rates, as in, e.g., \cite{jing-age-online,arafa-age-online-finite,bacinoglu-aoi-eh-finite-gnrl-pnlty}.} yet are subject to {\it erasures;} each transmission may get erased with probability $q$. Erasure events are independent and identically distributed (i.i.d.) across transmissions. The destination provides feedback to denote successful/failed transmissions. An overview of the system is shown in Fig.~\ref{fig:sys-mod}.

Let $l_{i,j}$ denote the $i$th transmission time pertaining to source $j$, and $s_{i,j}$ denote the $i$th {\it successful} one of which. Clearly, due to erasures,  $\{s_{i,j}\}\subseteq\{l_{i,j}\}$. Let us define $\mathcal{E}(t)$ and $\mathcal{D}(t)$ as the energy available in the sensor's battery and the identity of the data packet available in the sensor's data buffer at time $t$, respectively. Note that $\mathcal{E}(t)\in\{0,1\}$, while $\mathcal{D}(t)\in\{0,1,2,\dots,N\}$, with $\mathcal{D}(t)=0$ denoting an empty data buffer. Therefore, we have the following {\it energy causality} and {\it data causality} constraints: 
\begin{align}
\mathcal{E}\left(l_{i,j}^-\right)=1,~\mathcal{D}\left(l_{i,j}^-\right)=j, \quad \forall i,j, \label{eq:causality}
\end{align}
where $l_{i,j}^-$ denotes the time instant right before $l_{i,j}$. A set of feasible $\{l_{i,j}\}$ according to (\ref{eq:causality}) is denoted the {\it transmission policy}. We denote this by $\pi$ a {\it scheduling policy} that determines how the sensor manages its data buffer, e.g., which data source to be given priority. Observe that the transmission policy is, in general, highly intertwined with the scheduling policy.

\begin{figure}
\centering
\includegraphics[scale=.725]{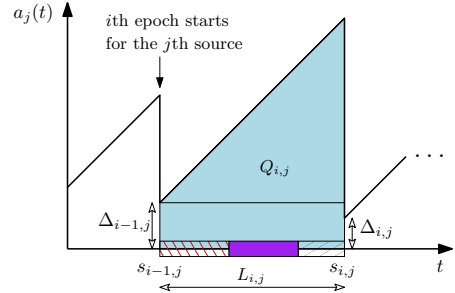}
\caption{An example evolution of the $j$th source AoI in the $i$th epoch. Falling (resp.~rising) hashed lines rectangles denote failed (resp.~successful) attempts for source $j$, while the solid rectangle denotes other sources' attempts.}
\label{fig:aoi_ex_source_j}
\vspace{-.2in}
\end{figure}

Our main metric of focus is data freshness, captured through AoI. When a transmission for source $j$'s data is successful, a {\it status update} is received at the destination. The AoI for source $j$ at time $t$ is defined as
\begin{align}
a_j(t)\triangleq t-u_j(t),
\end{align}
where $u_j(t)$ is the time-stamp of the latest successfully-received data pertaining to source $j$. An example of how the AoI may evolve over time is shown in Fig.~\ref{fig:aoi_ex_source_j}. We use the term {\it epoch} to denote the time in between two consecutive successful transmissions for a given source. For source $j$'s $i$th epoch, we denote its starting AoI by $\Delta_{i-1,j}$, its length by $L_{i,j}$, and the area under the AoI evolution curve during which by $Q_{i,j}$, see Fig.~\ref{fig:aoi_ex_source_j}. From the figure, one can see that
\begin{align}
L_{i,j}=s_{i,j}-s_{i-1,j},\quad Q_{i,j}=\Delta_{i-1,j}L_{i,j}+\frac{1}{2}L_{i,j}^2.
\end{align}
The goal is to design transmission and scheduling policies to minimize the {\it collective} long-term average AoI of all data sources. That is, to solve the following problem:
\begin{align} \label{opt:main-problem}
\min_{\{l_{i,j}\},~\pi}\quad\frac{1}{N}\sum_{j=1}^N\limsup_{n\rightarrow\infty}\frac{\sum_{i=1}^n\mathbb{E}\left[Q_{i,j}\right]}{\sum_{i=1}^n\mathbb{E}\left[L_{i,j}\right]} \quad \mbox{s.t.}\quad(\ref{eq:causality}),
\end{align}
where the expectation $\mathbb{E}\left[\cdot\right]$ is taken according to the underlying energy, data, and erasure distributions. We discuss the solution of problem (\ref{opt:main-problem}) over the next two sections, first for the single source case, followed by the multiple sources case.

\section{The Single Source Case} \label{sec:sgl_src}

In this section, we focus on problem (\ref{opt:main-problem}) for $N=1$ source, which will serve as a building block for $N\geq2$. In this case, no scheduling is needed, and hence we drop the index $j$. Since we aim at minimizing AoI, only the freshest data packet is kept at the sensor's data buffer, i.e., newly-generated data packets preempt old ones waiting for transmission, if any. 

\begin{figure}
\centering
\includegraphics[scale=.8]{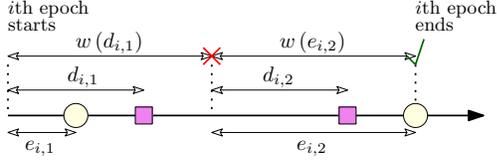}
\caption{An example for the $i$th epoch evolution with $M_i=2$. Circles (resp.~squares) represent energy (resp.~data) arrivals; a Cross (resp.~check mark) represents a failure (resp.~success).}
\label{fig:epoch_aoi_example}
\vspace{-.1in}
\end{figure}

Let $e_{i,1}$ and $d_{i,1}$ denote the time elapsed from the beginning of the $i$th epoch until the first energy and data arrivals, respectively. It then follows that $e_{i,1}\sim\exp\left(\lambda_e\right)$ and $d_{i,1}\sim\exp\left(\lambda_d\right)$. By (\ref{eq:causality}), the first transmission attempt in the $i$th epoch must therefore occur after at least $\max\{e_{i,1},d_{i,1}\}$ time units. Instead of transmitting right when energy and data are available, we allow the sensor to idly {\it wait} for some extra time units. While this lets the current data packet become more stale, {\it it provides an opportunity for the sensor to capture a fresher data packet in the waiting window before transmission.} Specifically, the first transmission attempt in the $i$th epoch occurs after
\begin{align}
w\left(\max\{e_{i,1},d_{i,1}\}\right)
\end{align}
time units from its beginning, for some waiting function $w(t)\geq t$. If such a transmission attempt fails, the above policy is repeated, yet with $e_{i,2}$ and $d_{i,2}$, which now denote the time until the next energy and data arrivals, respectively, {\it after} the first transmission attempt. By the memoryless property of the exponential distribution, $e_{i,2}\sim\exp\left(\lambda_e\right)$ and $d_{i,2}\sim\exp\left(\lambda_d\right)$ as well. Transmission attempts continue until success. Let $M_i$ denote the number of transmission attempts during the $i$th epoch. It is direct to see that $M_i$'s are i.i.d. geometrically-distributed with parameter $1-q$. Therefore, one can write
\begin{align}
L_i=\sum_{k=1}^{M_i}w\left(\max\{e_{i,k},d_{i,k}\}\right).
\end{align}
An example is shown in Fig.~\ref{fig:epoch_aoi_example}.

Next, observe that the proposed transmission/waiting policy induces a {\it stationary} distribution across all epochs. Since $w\left(\max\{e_{i,k},d_{i,k}\}\right)$'s are i.i.d., and since $M_i$ is independent of $w\left(\max\{e_{i,k},d_{i,k}\}\right)$, one can use Wald's identity to write
\begin{align}
\mathbb{E}\left[L_i\right]=&\mathbb{E}\left[M_i\right]\mathbb{E}\left[w\left(\max\{e_i,d_i\}\right)\right] \nonumber \\
=&\frac{\mathbb{E}\left[w\left(\max\{e_i,d_i\}\right)\right]}{1-q},\quad\forall i, \label{eq:exp_L_i}\\
\mathbb{E}\left[L_i^2\right]=&\mathbb{E}\left[M_i\right]\mathbb{E}\left[w\left(\max\{e_i,d_i\}\right)^2\right] \nonumber \\
&+\mathbb{E}\left[M_i(M_i-1)\right]\left(\mathbb{E}\left[w\left(\max\{e_i,d_i\}\right)\right]\right)^2 \nonumber \\
=&\frac{\mathbb{E}\left[w\left(\max\{e_i,d_i\}\right)^2\right]}{1-q} \nonumber \\
&+
\frac{2q\left(\mathbb{E}\left[w\left(\max\{e_i,d_i\}\right)\right]\right)^2}{(1-q)^2},\quad\forall i, \label{eq:exp_L_i_squared}
\end{align}
where the second equalities in (\ref{eq:exp_L_i}) and (\ref{eq:exp_L_i_squared}) follow from the properties of the geometric distribution. In addition, we note that $\Delta_{i-1}$ is now independent of $L_i$. Hence,
\begin{align}
\mathbb{E}\left[Q_i\right]=&\mathbb{E}\left[\Delta_{i-1}\right]\mathbb{E}\left[L_i\right]+\frac{1}{2}\mathbb{E}\left[L_i^2\right], \quad \forall i. \label{eq:exp_Q_i}
\end{align}
Using (\ref{eq:exp_L_i}), (\ref{eq:exp_L_i_squared}), and (\ref{eq:exp_Q_i}), problem (\ref{opt:main-problem}) reduces to the following optimization problem over a typical epoch:
\begin{align} \label{opt:epoch_problem}
\min_{w(t)\geq t}\quad\mathbb{E}\left[\Delta_{i-1}\right]+\frac{\mathbb{E}\left[L_i^2\right]}{2\mathbb{E}\left[L_i\right]}.
\end{align}

For a given waiting policy $w(\cdot)$, the following lemma characterizes the cumulative distribution function (CDF) of the starting AoI $\Delta_{i-1}$. By stationarity, we drop the index $i$ for simplicity of the presentation in the remainder of this section.

\begin{lemma}
The CDF of an epoch's starting AoI $\Delta$ is given by
\begin{align}
\!\!\!F_{\Delta}(\delta) = 1-e^{-\lambda_d \delta} P\left(w\left(\max\{e,d\}\right)-d\geq\delta\right),~\delta\geq0. \label{eq:CDF_Delta}
\end{align}
\end{lemma}

\begin{Proof}
We first note that $\Delta$ only depends on the variables pertaining to the successful (final) transmission attempt in the epoch, and does not depend on how many failures $M-1$ occurred before it. Thus, the random variables $e$ and $d$ denote the time until the energy and data arrivals, respectively, since the $(M-1)$th transmission attempt.

We now use total probability to write
\begin{align}
F_{\Delta}(\delta)=\int_{\substack{ t_e,t_d\geq0}}P\left(\Delta\leq\delta|e=t_e,d=t_d\right)f_{e,d}(t_e,t_d)dt_edt_d,
\end{align}
with $f_{e,d}(t_e,t_d)\triangleq\lambda_ee^{-\lambda_et_e}\lambda_de^{-\lambda_dt_d}$. Now observe that if $w\left(\max\{t_e,t_d\}\right)-t_d<\delta$, then clearly $P\left(\Delta\leq\delta|e=t_e,d=t_d\right)=1$. On the other hand, if $w\left(\max\{t_e,t_d\}\right)-t_d\geq\delta$, then $\Delta\leq\delta$ if and only if (iff) at least one data arrival occurred in the {\it last $\delta$ interval of the epoch.} The memoryless property of the exponential distribution indicates that $P\left(\Delta\leq\delta|e=t_e,d=t_d\right)=1-e^{-\lambda_d\delta}$ in this case. Combining both cases we get
\begin{align}
F_{\Delta}(\delta)=&1-e^{-\lambda\delta}\int_{t_e,t_d:~w\left(\max\{t_e,t_d\}\right)-t_d\geq\delta}\!\!f_{e,d}(t_e,t_d)dt_edt_d,
\end{align}
which is exactly (\ref{eq:CDF_Delta}).
\end{Proof}

We observe that solving problem (\ref{opt:epoch_problem}) is challenging since the waiting function is embedded into the CDF of $\Delta$ in a highly intertwined manner as shown in (\ref{eq:CDF_Delta}). Inspired by the results in \cite{jing-age-online, arafa-age-erasure-no-fb, arafa-age-erasure-fb, arafa-age-online-finite, bacinoglu-aoi-eh-finite-gnrl-pnlty} we focus on {\it threshold} waiting policies and analyze their performance. These are defined as
\begin{align} \label{eq:threshold_wait}
w(t)=t+\left[\gamma-t\right]^+,
\end{align}
for some $\gamma\geq0$, where $[\cdot]^+\triangleq\max(\cdot,0)$. Thus, a new transmission attempt takes effect only if the time until its pertaining energy and data become available surpasses a certain threshold $\gamma$. Threshold policies are quite intuitive, since one needs to balance the risk of waiting too much and letting the available data grow stale, with that of waiting too little and missing the opportunity to capture fresher data. In addition, they have been shown optimal in, e.g., \cite{jing-age-online, arafa-age-erasure-no-fb, arafa-age-erasure-fb, arafa-age-online-finite, bacinoglu-aoi-eh-finite-gnrl-pnlty}, albeit in a generate-at-will context where data arrival times are controlled. Under the $\gamma$-threshold policies in (\ref{eq:threshold_wait}), the next lemma characterizes the distribution of the starting AoI of each epoch. The proof is available in \cite{jrThesis} and is omitted here due to space limitations.

\begin{lemma}
Under a $\gamma$-threshold policy, the CDF of an epoch's starting AoI $\Delta$ is given by
\begin{align} \label{eq:CDF_Delta_gamma}
F_{\Delta}(\delta) \!=\! 1\!-\!e^{-\lambda_d\delta}&\bigg(1\!-\!e^{-\lambda_d\left[\gamma-\delta\right]^+} \nonumber \\
&\left.\times\left(1\!-\!\frac{\lambda_d}{\lambda_e+\lambda_d}e^{-\lambda_e\max\{\gamma,\delta\}}\right)\right),~\delta\geq0.
\end{align}
\end{lemma}

Using the CDF in (\ref{eq:CDF_Delta_gamma}), one can now compute the average starting AoI of the epoch as follows:
\begin{align}
\mathbb{E}\left[\Delta\right]=&\int_0^\infty\left(1-F_{\Delta}(\delta)\right)d\delta \nonumber \\
=&\frac{(1-e^{-\lambda_d\gamma})}{\lambda_d}-\gamma e^{-\lambda_d\gamma}\left(1-\frac{\lambda_d}{\lambda_e+\lambda_d}e^{-\lambda_e\gamma}\right) \nonumber \\
&+\frac{\lambda_d}{\left(\lambda_e+\lambda_d\right)^2}e^{-\left(\lambda_e+\lambda_d\right)\gamma}. \label{eq:exp_Delta}
\end{align}

Next, we evaluate the first and second moments of the epoch length $L$ in (\ref{eq:exp_L_i}) and (\ref{eq:exp_L_i_squared}), respectively, by evaluating the first and second moments of $w\left(\max\{e,d\}\right)$. Direct computations lead to the following result for the first moment:
\begin{align}
\mathbb{E}&\left[w\left(\max\{e,d\}\right)\right]=\gamma(1-e^{-\lambda_d \gamma})(1-e^{-\lambda_e \gamma}) 
\nonumber \\
&+\! \frac{(\lambda_d \gamma \!+\! 1)}{\lambda_d} e^{-\lambda_d \gamma} (1\!-\!e^{-\lambda_e \gamma}) 
+\! \frac{(\lambda_e \gamma \!+\! 1)}{\lambda_e} e^{-\lambda_e \gamma} (1\!-\!e^{-\lambda_d \gamma})
\nonumber \\
&+ \frac{\lambda_d [\lambda_e(\lambda_e + \lambda_d)\gamma + 2\lambda_e + \lambda_d]}{\lambda_e(\lambda_e + \lambda_d)^2}e^{-(\lambda_e + \lambda_d)\gamma} \nonumber \\
&+ \frac{\lambda_e [\lambda_d(\lambda_e + \lambda_d)\gamma + 2\lambda_d + \lambda_e]}{\lambda_d(\lambda_e + \lambda_d)^2}e^{-(\lambda_e + \lambda_d)\gamma}. \label{eq:exp_w}
\end{align}
For the second moment, the computations lead to a more involved expression, shown at the top of the next page in (\ref{eq:exp_w_squared}).

\begin{figure*}
\begin{align} \label{eq:exp_w_squared}
\mathbb{E}&\left[w\left(\max\{e,d\}\right)^2\right] =\gamma^2\left(1\!-\!e^{-\lambda_d \gamma}\right)\left(1\!-\!e^{-\lambda_e \gamma}\right) 
+ \frac{\lambda_d^2\gamma^2 \!+\! 2\lambda_d\gamma \!+\! 2}{\lambda_d^2}e^{-\lambda_d\gamma}\left(1 \!-\! e^{-\lambda_e\gamma}\right) 
+ \frac{\lambda_e^2\gamma^2 \!+\! 2\lambda_e\gamma \!+\! 2}{\lambda_e^2}e^{-\lambda_e\gamma}\left(1 \!-\! e^{-\lambda_d\gamma}\right)
\nonumber \\
&+ \frac{\lambda_d\left(\lambda_e^2(\lambda_e + \lambda_d)^2\gamma^2 + 2\lambda_e(\lambda_e + \lambda_d)(2\lambda_e + \lambda_d)\gamma + 6\lambda_e^2 + 6\lambda_e\lambda_d + 2\lambda_d^2\right)}{\lambda_e^2(\lambda_e + \lambda_d)^3}e^{-(\lambda_e + \lambda_d)\gamma}
\nonumber \\
&+ \frac{\lambda_e\left(\lambda_d^2(\lambda_e + \lambda_d)^2\gamma^2 + 2\lambda_d(\lambda_e + \lambda_d)(2\lambda_d + \lambda_e)\gamma + 6\lambda_d^2 + 6\lambda_e\lambda_d + 2\lambda_e^2\right)}{\lambda_d^2(\lambda_e + \lambda_d)^3}e^{-(\lambda_e + \lambda_d)\gamma}.
\end{align}
\hrulefill
\end{figure*}

Finally, using (\ref{eq:exp_Delta}), (\ref{eq:exp_w}), and (\ref{eq:exp_w_squared}), together with (\ref{eq:exp_L_i}) and (\ref{eq:exp_L_i_squared}), one can substitute in (\ref{opt:epoch_problem}) and evaluate the long-term average AoI achieved with a $\gamma$-threshold policy. We define this as $\overline{\texttt{AoI}}_q\left(\gamma\right)$ to emphasize the dependency on $q$ and $\gamma$. 




\section{The Multiple Sources Case}

In this section, we extend the results of Section~\ref{sec:sgl_src} to $N\geq2$ sources. We consider a {\it maximum-age-first} (MAF) scheduling policy, denoted $\pi_{MAF}$, in which the sensor's data buffer accepts data packets from source $j$ at time $t$ iff it has the maximum instantaneous AoI, i.e., iff $a_j(t)\geq a_{\kappa}(t),~\forall\kappa\neq j$. Let us assume without loss of generality that the system starts with fresh information at time $0$: $a_j(0)=0,~\forall j$, and hence, under $\pi_{MAF}$ the sensor first dedicates all transmission attempts to source $1$'s data, until successful, and then focuses on source $2$'s data, all the way until source $N$'s data is transmitted successfully, and then repeats transmission attempts in the same order $\{1,2,\dots,N\}$.\footnote{We note that MAF scheduling is only possible due to the erasure status feedback made available by the destination.}


Let us focus on some source $j$, and denote by $e^{(j)}_{i,1}$ and $d^{(j)}_{i,1}$ the time elapsed from the beginning of the $i$th epoch until the first energy and data arrival, respectively, dedicated for that source. As in Section~\ref{sec:sgl_src}, the sensor does not immediately attempt transmission after receiving the energy and data. Instead, the first transmission attempt for source $j$ in the $i$th epoch occurs after
\begin{align}
w\left(\max\left\{e^{(j)}_{i,1},d^{(j)}_{i,1}\right\}\right)
\end{align}
time units from its beginning. This is followed by a second attempt in case of failure, which occurs after another $w\left(\max\left\{e^{(j)}_{i,2},d^{(j)}_{i,2}\right\}\right)$ time units, where $e^{(j)}_{i,2}$ and $d^{(j)}_{i,2}$ now denote the time until the next energy and data arrivals, respectively, for source $j$ {\it after} the first transmission attempt. This continues until source $j$'s transmission is successful, which takes $M_{j,i}$ attempts. Afterwards, the focus turns to source $j+1$. 

Observe that $M_{j,i}$'s are i.i.d. geometric random variables with parameter $1-q$. In addition, by the memoryless property of exponential distribution, $e^{(j)}_{i,k}\sim\exp\left(\lambda_e\right)$ and $d^{(j)}_{i,k}\sim\exp\left(\lambda_{d,j}\right)$, $\forall i,k$. The structure of our waiting policy, therefore, induces a stationary distribution across all epochs. Therefore, we drop the index $i$, and define the following random variables in a typical epoch for source $j$: $\Delta^{(j)}$ as the starting AoI; $L^{(j)}$ as the epoch length; and $Q^{(j)}$ as the area under the AoI evolution curve in the epoch. Therefore, one can write
\begin{align} \label{eq:Q_j_multi}
\mathbb{E}\left[Q^{(j)}\right]=\mathbb{E}\left[\Delta^{(j)}\right]\mathbb{E}\left[L^{(j)}\right]+\frac{1}{2}\mathbb{E}\left[\left(L^{(j)}\right)^2\right].
\end{align}

As in Section~\ref{sec:sgl_src}, we focus on $\gamma$-threshold waiting policies in our analysis. We use the same threshold $\gamma$ for all sources.\footnote{The analysis is readily extendable to account for different thresholds.} Now observe that under $\pi_{MAF}$, source $j$'s epoch length depends on the time elapsed until {\it all other sources are done with their successful transmissions.} With a slight abuse of notation, let us denote by $L_{\kappa}$ the time needed for source $\kappa$ to finish its successful transmission. Therefore, one can express
\begin{align} \label{eq:L_j_multi}
L^{(j)}=\sum_{\kappa=1}^NL_{\kappa}
\end{align}
in a typical epoch. We now present the main result. 

\begin{theorem}
Let $\overline{\texttt{AoI}}_{q,N}\left(MAF,\gamma\right)$ denote the collective long-term average AoI of problem (\ref{opt:main-problem}) achieved under $\pi_{MAF}$ and $\gamma$-threshold waiting policy. Then
\begin{align}
&\overline{\texttt{AoI}}_{q,N}\left(MAF,\gamma\right) \nonumber \\
&= \frac{1}{N}\sum_{j=1}^N \mathbb{E}\left[\Delta^{(j)}\right] + \frac{\sum_{\kappa=1}^N\mathbb{E}\left[w\left(\max\{e^{(\kappa)},d^{(\kappa)}\}\right)^2\right]}{2 \sum_{\kappa=1}^N\mathbb{E}\left[w\left(\max\{e^{(\kappa)},d^{(\kappa)}\}\right)\right]} 
\nonumber \\
&+ \frac{q\sum_{\kappa=1}^N\left(\mathbb{E}\left[w\left(\max\{e^{(\kappa)},d^{(\kappa)}\}\right)\right]\right)^2}{(1-q) \sum_{\kappa=1}^N\mathbb{E}\left[w\left(\max\{e^{(\kappa)},d^{(\kappa)}\}\right)\right]} \nonumber \\ 
&+\! \frac{\sum_{\substack{1\leq\alpha\leq N \\ 1\leq\beta<\alpha}} \mathbb{E}\!\left[w\left(\max\{e^{(\alpha)}\!,\!d^{(\alpha)}\}\right)\right] \! \mathbb{E}\!\left[w\left(\max\{e^{(\beta)}\!,\!d^{(\beta)}\}\right)\right]}{(1-q)\sum_{\kappa=1}^N\mathbb{E}\left[w\left(\max\{e^{(\kappa)},d^{(\kappa)}\}\right)\right]},
\label{eq:Delta_MAF}
\end{align}
with $\mathbb{E}\left[\Delta^{(j)}\right]$ given by (\ref{eq:exp_Delta}) after replacing $\lambda_d$ with $\lambda_{d,j}$, and the first and second moments of $w\left(\max\{e^{(\kappa)},d^{(\kappa)}\}\right)$ given by (\ref{eq:exp_w}) and (\ref{eq:exp_w_squared}), respectively, after replacing $\lambda_d$ with $\lambda_{d,\kappa}$.
\end{theorem}

\begin{Proof}
It is clear from (\ref{eq:L_j_multi}) that $L^{(j)}$'s are i.i.d. across sources $\sim L^{(\star)}$. By (\ref{eq:Q_j_multi}), one can express $\overline{\texttt{AoI}}_{q,N}\left(MAF,\gamma\right)$ as
\begin{align} \label{eq:Delta_MAF_proof}
\frac{1}{N}\sum_{j=1}^N\mathbb{E}\left[\Delta^{(j)}\right]+\frac{\mathbb{E}\left[\left(L^{(\star)}\right)^2\right]}{2\mathbb{E}\left[L^{(\star)}\right]}.
\end{align}
Now observe that the average starting AoI $\mathbb{E}\left[\Delta^{(j)}\right]$ will be given by (\ref{eq:exp_Delta}) after replacing $\lambda_d$ by $\lambda_{d,j}$, since the same $\gamma$-threshold policy is applied at every transmission attempt. Thus, it only remains to evaluate the first and second moments of $L^{(\star)}$. Towards that end, using (\ref{eq:L_j_multi}), one can write
\begin{align}
\mathbb{E}\left[L^{(\star)}\right]&=\sum_{\kappa=1}^N\mathbb{E}\left[L_{\kappa}\right], \\
\mathbb{E}\left[\left(L^{(\star)}\right)^2\right]&=\sum_{\kappa=1}^N\mathbb{E}\left[\left(L_{\kappa}\right)^2\right]+2\sum_{\alpha=1}^N\sum_{\beta=1}^{\alpha-1}\mathbb{E}\left[L_{\alpha}\right]\mathbb{E}\left[L_{\beta}\right].
\end{align}
Next, we note that the first and second moments of $L_\kappa$ are given by (\ref{eq:exp_L_i}) and (\ref{eq:exp_L_i_squared}), respectively, in which the corresponding first and second moments of the waiting random variables are given by (\ref{eq:exp_w}) and (\ref{eq:exp_w_squared}), respectively, after replacing $\lambda_d$ with $\lambda_{d,\kappa}$. Substituting these in (\ref{eq:Delta_MAF_proof}) and simplifying gives (\ref{eq:Delta_MAF}).
\end{Proof}

\section{Numerical Evaluations} \label{sec:num_rslt}

\begin{figure}[t]
\centering
\includegraphics[scale=.4]{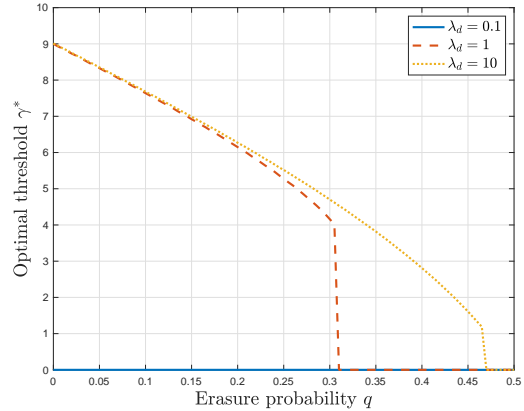}
\caption{Optimal threshold $\gamma^*$ versus the erasure probability $q$, with $\lambda_e=0.1$ and varying values of $\lambda_d$.}
\label{fig:opt_gamma_e_pt1}
\vspace{-.1in}
\end{figure}

We now present various numerical evaluations to further illustrate the results of this paper. We first show how the optimal threshold value behaves as a function of the system parameters. In all experiments, we set the energy arrival rate to $\lambda_e=0.1$. In Fig.~\ref{fig:opt_gamma_e_pt1}, we plot the optimal threshold $\gamma^*$ that minimizes $\overline{\texttt{AoI}}_q\left(\gamma\right)$ versus the erasure probability $q$, with varying values of $\lambda_d$. One can see that as the erasure probability increases, the optimal threshold value decreases, which demonstrates that waiting for additional data to arrive is not beneficial for the AoI due to the increased rate of erased data transmissions. We also observe that as the data arrival rate approaches the energy arrival rate, the optimal threshold value decreases. This shows that {\it waiting is more beneficial to reducing the AoI when $\lambda_d$ is relatively larger than $\lambda_e$, and when $q$ is relatively small.}

\begin{figure}[t]
\centering
\includegraphics[scale=.4]{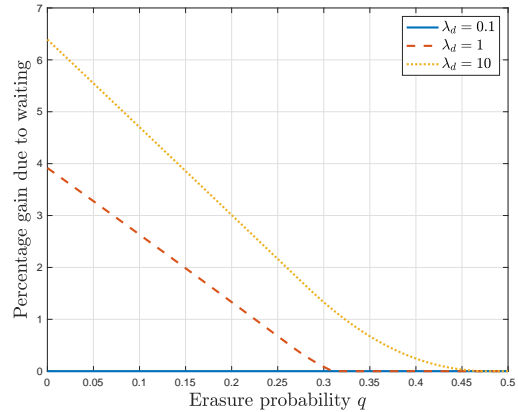}
\caption{Percentage gain due to waiting versus the erasure probability, with $\lambda_e=0.1$ and varying values of $\lambda_d$.}
\label{fig:gain_waiting_e_pt1}
\vspace{-.1in}
\end{figure}

In Fig.~\ref{fig:gain_waiting_e_pt1}, we plot the {\it percentage gain due to waiting} versus the erasure probability. We define the percentage gain as
\begin{align}
\left(1-\frac{\overline{\texttt{AoI}}_q\left(\gamma^*\right)}{\overline{\texttt{AoI}}_q\left(0\right)}\right)\times100\%.
\end{align}
That is, the percentage amount of {\it reward} one can gain by applying the optimal threshold waiting policy when compared to a {\it zero-wait} policy. From the figure, it can be seen that as the data arrival rate approaches the energy arrival rate, there is no percentage gain to waiting for additional data arrivals. However, as the data arrival rate becomes relatively larger than the energy arrival rate, waiting becomes significantly beneficial with the corresponding optimal threshold value shown in Fig.~\ref{fig:opt_gamma_e_pt1}. In addition, since $\gamma^*$ approaches $0$ as $q$ increases, we see that the percentage gain due to waiting decreases with the increase in erasure probability as well. Finally, though it is not shown on the figure, we observe, numerically, that for $\lambda_d>10$, the percentage gain curve is almost the same as that for $\lambda_d=10$. This may be attributed to the fact that the sensor's battery is unit-sized, and therefore higher gains from waiting could be achieved for larger battery sizes.

\begin{figure}[t]
\centering
\includegraphics[scale=.4]{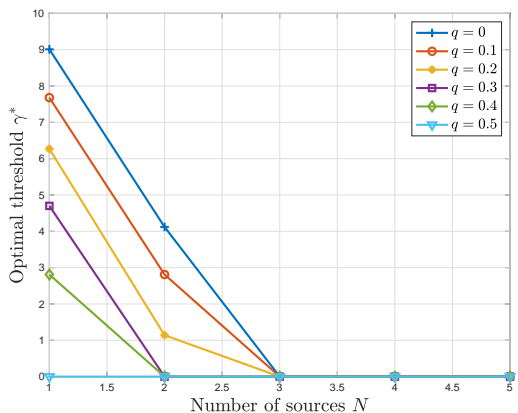}
\caption{Optimal threshold $\gamma^*$ versus the number of sources $N$, with $\lambda_e=0.1$, $\lambda_{d,j}=10,~\forall j$, and varying values of $q$.}
\label{fig:opt_gamma_multi_e_pt1_d_10}
\vspace{-.1in}
\end{figure}

Next, we present results for the multiple sources case. For that, we focus on a {\it symmetric} system in which all data arrivals' rates are the same, i.e., $\lambda_{d,j}=\lambda_d,~\forall j$. Hence, we drop the sources' indices from (\ref{eq:Delta_MAF}), since every random variable now is identical, and simplify the expression to get   
\begin{align}
\overline{\texttt{AoI}}_{q,N}\left(MAF,\gamma\right)&=\mathbb{E}\left[\Delta\right]+\frac{\mathbb{E}\left[\left(w\left(\max\{e,d\}\right)\right)^2\right]}{2\mathbb{E}\left[w\left(\max\{e,d\}\right)\right]} \nonumber \\
&+\left(\frac{q+\frac{N-1}{2}}{1-q}\right)\mathbb{E}\left[w\left(\max\{e,d\}\right)\right].
\end{align}
It is immediate to see that the collective long-term average AoI is increasing in both the number of sources $N$ and the erasure probability $q$. In Fig.~\ref{fig:opt_gamma_multi_e_pt1_d_10}, we show how the optimal threshold $\gamma^*$ behaves as a function of $N$ and $q$. Fig.~\ref{fig:opt_gamma_multi_e_pt1_d_10} demonstrates that as the number of sources grow relatively large, there is no benefit to waiting for additional data arrivals and the corresponding optimal threshold policy becomes a zero-wait policy. This is mainly because as the number of sources increase, each source's inter-update duration becomes {\it longer}, since they need to wait for each other under the $\pi_{MAF}$ policy. It is also shown in Fig.~\ref{fig:opt_gamma_multi_e_pt1_d_10}, as in Figs.~\ref{fig:opt_gamma_e_pt1} and~\ref{fig:gain_waiting_e_pt1}, that the optimal threshold value decreases as a function of $q$. Once $q = 0.5$, the optimal threshold values become $0$ for any number of sources, which agrees with the data shown in Figs.~\ref{fig:opt_gamma_e_pt1} and~\ref{fig:gain_waiting_e_pt1}. This also resonates with the results shown in the generate-at-will single source study of \cite{arafa-age-erasure-no-fb}, in which zero-waiting is optimal if $q\geq0.5$.

\section{Conclusions}

A multiple source status updating system has been considered, in which data is generated according to Poisson processes and are conveyed to a destination over an erasure channel using a shared energy harvesting sensor. Detailed analyses of the achievable collective long-term average AoI of the sources have been carried out with a focus on threshold-based transmission policies combined with maximum-age-first scheduling, showcasing the benefits of waiting before updating in such systems and extending previous works in the literature.

Future work includes analyzing other scheduling policies for sensors with larger data buffers and battery sizes.

\end{document}